# Predictive Dynamic Scheduling for Deterministic Communications in Beyond 5G

Syed Morsleen Riaz, M. Carmen Lucas-Estañ, Baldomero Coll-Perales, Javier Gozalvez
*Uwicore laboratory, Universidad Miguel Hernández de Elche, Elche (Alicante), Spain*
sriaz@umh.es, m.lucas@umh.es, bcoll@umh.es, j.gozalvez@umh.es

***Abstract*—Next generation wireless networks must sustain deterministic service levels to support emerging time-sensitive applications. The ability to guarantee bounded latencies depends on the efficient management of radio resources. Several studies propose leveraging the native intelligence of future networks to develop predictive schedulers capable of efficiently managing resources. However, existing proposals focus on semi-static scheduling, where resources are reserved based on traffic predictions, and these reservations are susceptible to inefficiencies due to prediction inaccuracies. This study advances the state of the art with a novel predictive dynamic scheduling scheme that avoids such inefficiencies, and leverages traffic predictions to allocate resources to incoming requests that meet their latency requirements while avoiding resources likely to be needed by future predicted packets. Our results demonstrate that the proposed predictive dynamic scheduling effectively supports deterministic communications in scenarios with mixed traffic flows and varying QoS requirements.**

## I. Introduction

Future networks must support emerging time-sensitive and deterministic services driven by the digitalization of key verticals such as connected and autonomous mobility and Industry 4.0. Delivering deterministic wireless communications with bounded (and low) latency and high reliability in beyond-5G networks remains a challenge due to the inherently stochastic nature of wireless channels and systems. This challenge becomes even more complex when managing mixed traffic flows with diverse requirements, including time-sensitive and deterministic traffic.

The ability to support deterministic services in mixed traffic flows heavily depends on efficient radio resource management and the design of scheduling mechanisms capable of meeting diverse Quality of Service (QoS) requirements. In 5G NR, scheduling can be either dynamic or semi-static [1]. With dynamic scheduling, radio resources are allocated on a per-packet basis upon receiving a scheduling request. In contrast, semi-static scheduling - configured grants for uplink and semi-persistent scheduling for downlink- pre-allocates radio resources periodically to users. By eliminating the need for scheduling requests (SRs) and subsequent grants before data transmission, semi-static scheduling reduces latency compared to dynamic scheduling. However, when the packet arrival frequency does not align with the periodicity of resource allocation, or when packet sizes vary, semi-static scheduling can lead to inefficient resource utilization or failure to meet latency and determinism requirements [2]. One way to address this challenge is by assigning and managing multiple configured grants per user with different periodicities [3]. However, this approach may result in resource overprovisioning to accommodate the most stringent QoS requirements or increased signaling overhead to dynamically activate the most suitable configured grant [4]. Other alternatives to improve resource utilization and the likelihood of meeting deterministic requirements include, for example, the use of Satisfiability Modulo Theory (SMT) [5] to identify feasible resource allocation configurations for 5G configured grant scheduling of real-time industrial traffic. Preemption can also be used to prioritize certain traffic (e.g., deterministic) but this is done at the cost of interrupting lower-priority transmissions (e.g., best-effort), which can lead to increased latency and packet losses for the latter [6]. Despite these advancements, the lack of knowledge about future network conditions and upcoming traffic demands, combined with the presence of mixed traffic with diverse communication requirements, makes it challenging for 5G schedulers to efficiently allocate resources while ensuring the deterministic requirements of time-sensitive traffic [6].

Several studies propose leveraging the native AI/ML capabilities of future networks to develop predictive schedulers that efficiently manage radio resources and improve the capacity to support services with stringent QoS requirements using traffic demand forecasts ([7]-[10]). However, existing proposals focus on semi-static scheduling, where resources are pre-allocated or reserved based on traffic predictions. Such pre-allocations are susceptible to resource allocation inefficiencies due to inaccuracies in predictions or the stochastic nature of wireless systems and traffic sources, which may reduce the capacity to guarantee the determinism required for time-sensitive communications [11]. In this context, this paper advances the state-of-the-art with a novel predictive dynamic scheduling scheme that avoids possible inefficient resource reservations caused by inaccurate predictions. Unlike semi-static schedulers, the proposed predictive dynamic scheduling scheme allocates resources only upon receiving a scheduling request. However, it leverages traffic predictions to prioritize allocations that meet the current transmission's requirements while avoiding resources likely to be needed by future predicted packets. This study demonstrates how the proposed predictive dynamic scheduling scheme effectively utilizes traffic predictions and varying QoS requirements in mixed traffic flows to enhance support for deterministic communications in beyond-5G.

The rest of the paper is organized as follows. Section II reviews the state-of-the-art, and Section III presents the proposed predictive dynamic scheduling scheme. Section IV details the traffic characterization and prediction processes implemented in this study, and Section V compares the performance of the predictive dynamic scheduler against standard 5G dynamic scheduling. Finally, Section VI summarizes the main conclusions of this study.

This work has been partially funded by the European Commission Horizon Europe SNS JU 6G-SHINE (GA 101095738) project, and by MCIN/AEI/10.13039/501100011033 (PID2020-115576RB-I00, PID2023-150308OB-I00).

## II. State Of The Art

Predictive solutions will play a key role in next-generation networks since they can improve radio resource utilization efficiency and reduce excessive over-dimensioning when supporting traffic with stringent requirements, including the deterministic requirements of time-sensitive services [12]. Predictive schedulers have emerged as a key approach for managing Time-Sensitive Networks (TSN). In [13], the authors introduce an online traffic scheduler that leverages deep reinforcement learning (DRL) and convolutional neural networks (CNN) to extract flow features and optimize scheduling and resource allocation in TSN. [14] proposes a deterministic federated learning framework and a DRL-based resource scheduling algorithm for managing time-sensitive industrial IoT services. [15] expands the use of predictive schedulers and explores their application to manage traffic flows with different QoS profiles (or 5G QoS Identifiers, 5QIs) in a 5G asynchronous deterministic backhaul network. The authors propose a reinforcement learning-based flow scheduler that utilizes predictive data analytics – such as flow lifetime duration, packet arrival rate, and delay budget statistics- to increase the number of supported flows in deterministic asynchronous networks. The study demonstrates that predictive schedulers can better meet the requirements of deterministic traffic by proactively managing resources based on traffic predictions.

Several studies also propose predictive schedulers to efficiently manage radio resources at the RAN (Radio Access Networks) and better support services using traffic demand forecasts. For example, [7] proposes a semi-persistent downlink scheduler integrated with a short-term traffic predictor. The authors show that the predictive scheduler achieves performance comparable to traditional proportional-fair dynamic scheduler while reducing computational complexity in the presence of bursty video traffic. In [8], the authors propose a DRL grant allocator combining offline and online learning for supporting URLLC (Ultra-Reliable Low Latency Communication) traffic with uplink semi-static grant-free schedulers. Grant-free scheduling enables data transmissions over pre-allocated resources, reducing signaling overhead and communication latency. Additionally, [9] develops a multi-objective DRL technique for priority-enabled grant-free (GF) or configured grant (CG) scheduling of uplink traffic in a massive Machine-Type Communication (mMTC) scenario, where heterogeneous MTC devices transmit small data packets and compete for shared GF resources. The authors demonstrate that predictive schedulers improve the probability of successful transmissions by reducing collisions in shared resource access. [10] presents an alternative to using traffic predictors in predictive schedulers. The authors analyze correlations between process activations in industrial environments and data generation patterns to identify spatio-temporal traffic correlations. They then exploit these correlations to optimize semi-static scheduling configurations in beyond-5G networks.

Existing proposals highlight the potential of predictive schedulers to support services with stringent QoS requirements. However, all current approaches focus on semi-static scheduling, where resources are reserved based on traffic predictions, and these reservations remain susceptible to resource management inefficiencies due to inaccuracies in predictions or the stochastic nature of wireless systems and traffic sources. To overcome these limitations, we propose a predictive dynamic scheduler that leverages traffic predictions to select resource allocations that can satisfy the demand of current transmissions and increase the likelihood of satisfying future transmissions.

## III. Predictive Dynamic Scheduling

This study introduces the first Predictive Dynamic Scheduling (PDS) scheme for Beyond 5G. PDS leverages traffic predictions to prioritize resource allocations that meet the requirements for transmissions that sent a Scheduling Request (SR) while avoiding resources likely to be needed by future predicted packets. Before presenting our PDS proposal, we first highlight the potential of PDS in scenarios with mixed traffic flows and varying QoS requirements.

### A. *On the potential for predictive dynamic scheduling*

In 5G NR, dynamic scheduling allocates radio resources on a per-packet basis upon receiving a SR. The scheduler can implement various policies to select among the available radio resources based on the SR. Fig. 1 illustrates an example where a new SR for the transmission of packet $pkt_0$ is generated at $t_0$. This packet must be transmitted within a maximum latency deadline $d_0$ and requires $R_0$=9 radio resources. In NR, a radio resource is defined by a Resource Block (RB) in the frequency domain and a slot in the time domain. The RBs must be selected within the transmission window to meet the latency deadline. The transmission window includes all available RBs between $t_0$ and $d_0$. Fig. 1-left shows that two allocation options within this window would meet the latency deadline. A common policy is to minimize latency, selecting option 1 in Fig. 1-left. Now, suppose that a new SR for packet $pkt_1$, requiring $R_1$=6 RBs with a latency deadline $d_1$, arrives at $t_1$ (Fig. 1-middle). The figure shows that it is not possible to meet $pkt_1$'s latency deadline as the necessary RBs within its transmission window were previously assigned to $pkt_0$. However, both packets could meet their latency requirements if the scheduler had anticipated at $t_0$ that a new SR for $pkt_1$ (demanding $R_1$=6 RBs) would be received at $t_1$. In this case, the scheduler could have assigned to $pkt_0$ the RBs corresponding to option 2 in Fig. 1-left, allowing $pkt_1$ to receive the necessary RBs and meet its latency deadline $d_1$ (Fig. 1-right). This example highlights the potential of predictive dynamic scheduling, and the performance gains it can achieve in scenarios with mixed traffic flows and varying QoS requirements.

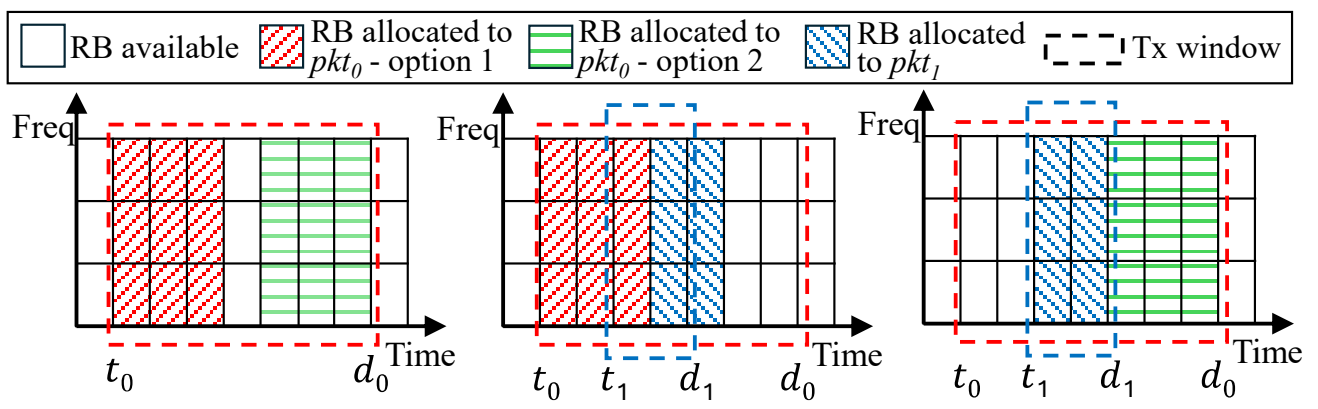

Fig. 1. On the potential of predictive dynamic scheduling.

### B. *Predictive Dynamic Scheduler*

Consider a 5G NR cell supporting $N$ nodes generating data for uplink transmission. Let $pkt_0$ be a packet of size $s_0$ generated by a node $n$ at time $t_0$ with a latency requirement $L_0$. The node $n$ sends an SR to the gNB to request RBs for transmitting this packet. The gNB determines the number of RBs as $R_0 = f(s_0, mcs_0)$, where $f(\cdot)$ is a function that determines the number of RBs needed to transmit a packet of size $s_0$ using a Modulation and Coding Scheme (MCS) $mcs_0$ following [1]. A standard dynamic scheduler searches for $R_0$ RBs available in consecutive slots within its transmission window, i.e., between the packet's generation time $t_i$ and its deadline $d_i = t_i + L_i$. In contrast, the PDS proposal searches for the $R_0$ RBs considering not only the requirements of $pkt_0$ but also those of future predicted packets $\widehat{pkt}_i$. To this aim, a predictor forecasts the generation time $\hat{t}_i$ and size $\hat{s}_i$ of the next $P$ packets after $pkt_0$. The scheduler then estimates the number of RBs required for each of these packets as $\widehat{R}_i = f(\hat{s}_i, mcs_i)$, with each packet to be transmitted before $\widehat{d}_i = \hat{t}_i + L_i$.

PDS avoids assigning $pkt_0$ RBs that may be needed for future packets, ensuring that latency requirements are met for both current and upcoming packets. To this end, PDS identifies first the transmission window of $pkt_0$ and the available RBs within the window (Fig. 2.a). PDS then identifies the set Φ of predicted packets whose transmission window may overlap with that of $pkt_0$. Two packets $pkt_i$ and $pkt_j$ have overlapping transmission windows if $t_i < d_j$ and $t_j \leq d_i$. In Fig. 2.b, the transmission windows of $\widehat{pkt}_1$, $\widehat{pkt}_2$, and $\widehat{pkt}_3$ overlap with that of $pkt_0$, while non-overlapping windows (e.g. $\widehat{pkt}_4$) do not affect the transmission of $pkt_0$ and are not considered by PDS. PDS temporarily marks as *RBs to avoid* the available RBs between $\hat{t}_i$ and $\widehat{d}_i$ for each predicted packet $\widehat{pkt}_i \subset \Phi$ as shown in Fig. 2.c. Even if the number of RBs within $\hat{t}_i$ and $\widehat{d}_i$ exceeds the $\widehat{R}_i$ RBs required by $\widehat{pkt}_i$ (computed based on $\hat{s}_i$), PDS reserves them to be able to accommodate upcoming packets even with inaccurate $\hat{t}_i$ and $\hat{s}_i$ predictions. PDS tries to avoid allocating $pkt_0$ any RBs between $\hat{t}_i$ and $\widehat{d}_i$ for all $\widehat{pkt}_i$ ($i = 1, 2,\ldots, P$) to ensure $pkt_i$ will have sufficient RBs available to meet its latency deadline.

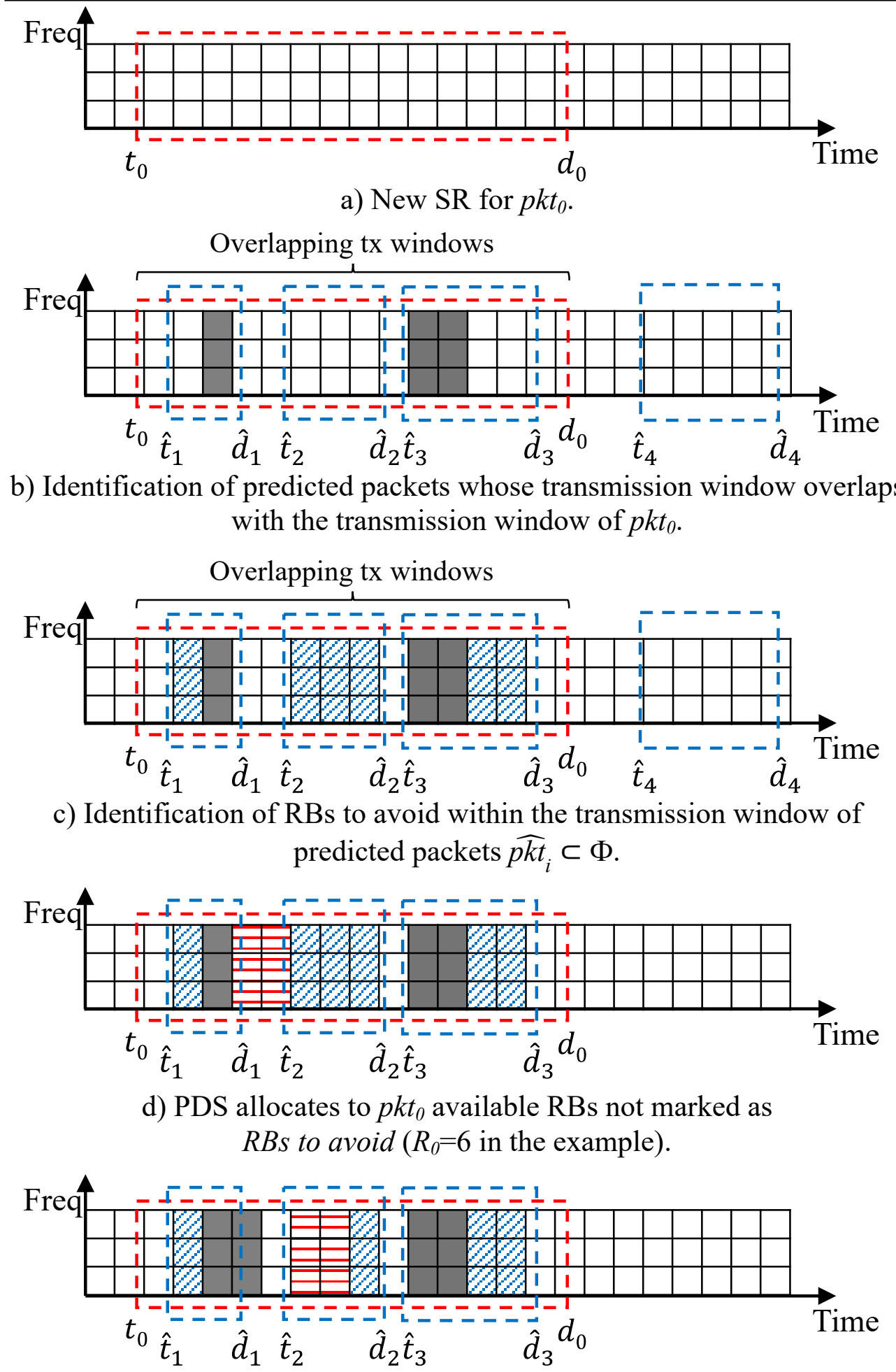

a) New SR for $pkt_0$.

b) Identification of predicted packets whose transmission window overlaps with the transmission window of $pkt_0$.

c) Identification of RBs to avoid within the transmission window of predicted packets $\widehat{pkt}_i \subset \Phi$.

d) PDS allocates to $pkt_0$ available RBs not marked as *RBs to avoid* ($R_0$=6 in the example).

e) PDS allocates to $pkt_0$ RBs initially marked as *RBs to avoid* within the transmission window of $\widehat{pkt}_2$.

Fig. 2. Illustration of the operation of the PDS proposal.

PDS then searches for $R_0$ available RBs in consecutive slots for the transmission of $pkt_0$. These RBs must be within the transmission window of $pkt_0$ to meet its latency deadline (Fig. 2.d). If multiple options are available, PDS selects the RBs that offer the lowest latency for $pkt_0$. If no viable options exist, PDS must allocate to $pkt_0$ RBs initially marked as *RB to avoid*. This situation is illustrated in Fig. 2.e where the number of unavailable RBs (allocated to previous transmissions) has been modified. PDS identifies the predicted packet $\widehat{pkt}_i \subset \Phi$ with the largest latency requirement ($\widehat{pkt}_3$ in Fig. 2) and checks whether the number of available RBs within its transmission window $\widehat{U}_i$ (between $\hat{t}_i$ and $\widehat{d}_i$) is equal to or greater than the sum of RBs required to transmit $pkt_0$ and $\widehat{pkt}_i$, i.e. $\widehat{U}_i \geq R_0 + \widehat{R}_i$. If this condition is satisfied, PDS allocates $R_0$ RBs for $pkt_0$ in the first available slots after $\hat{t}_i$. Otherwise, PDS evaluates the same condition for $\widehat{pkt}_j \subset \Phi$ with the next largest latency requirement $L_i$. In the example in Fig. 2.e, we consider $R_0$=6, $\widehat{R}_2$=3, and $\widehat{R}_3$=3. The transmission window of $\widehat{pkt}_3$ has only 6 available RBs, which is insufficient to accommodate both $pkt_0$ and $\widehat{pkt}_3$. PDS then checks whether the available RBs within the transmission window of $\widehat{pkt}_2$ are sufficient to accommodate $pkt_0$ and $\widehat{pkt}_2$. Since $\widehat{U}_2$= 9 in Fig. 2.e, which satisfies the condition $\widehat{U}_2 \geq R_0$+$\widehat{R}_2$, PDS allocates RBs to $pkt_0$ within the transmission window of $\widehat{pkt}_2$. If the condition $\widehat{U}_i \geq R_0 + \widehat{R}_i$ cannot be met for any predicted packets in Φ, PDS allocates $R_0$ RBs to $pkt_0$ from the available RBs within the transmission window $\widehat{U}_i$ of the predicted packet $\widehat{pkt}_i \subset \Phi$ with the largest latency requirement ($\widehat{pkt}_3$ in Fig. 2).

## IV. Traffic Characterization and Prediction

The proposed PDS scheme leverages predictive knowledge of future traffic to schedule current transmissions while accounting for upcoming traffic demands. To evaluate our proposal, we consider a 6G-envisioned autonomous driving scenario in which data collected or generated by an autonomous vehicle is sent to the network for processing at the edge [16]. Communications must meet a bounded latency deadline to ensure that offloading processing workloads to the network does not disrupt the vehicle's operation. For our evaluation, we use realistic sensor data generated by an autonomous vehicle through the Connected and Automated Mobility (CAM) platform presented in [17]. This platform integrates realistic sensing and autonomous driving (AD) capabilities using the open-source CARLA and AUTOWARE softwares, interconnected via a Robot Operative System (ROS) bridge. The platform generates sensor data from an autonomous vehicle with a full suite of Level 3 (L3) AD sensors, including five cameras and five radars mounted on the top, front, rear and sides of the vehicle. These sensors detect objects such as vehicles, obstacles and pedestrians, which are then processed by the AD software to control the vehicle. The sensors generate raw data at periodic intervals, which is then processed by a perception module to extract detected objects. The processed sensor data is transmitted to the network for processing at the edge. We have collected extensive datasets of the processed sensor packets -including size, timestamp and sensor ID- from realistic urban environments. Fig. 3 shows an example of processed sensor data packets generated with the CAM platform. The packet size varies over time based on the number of objects detected by each sensor as the vehicle navigates an urban scenario.

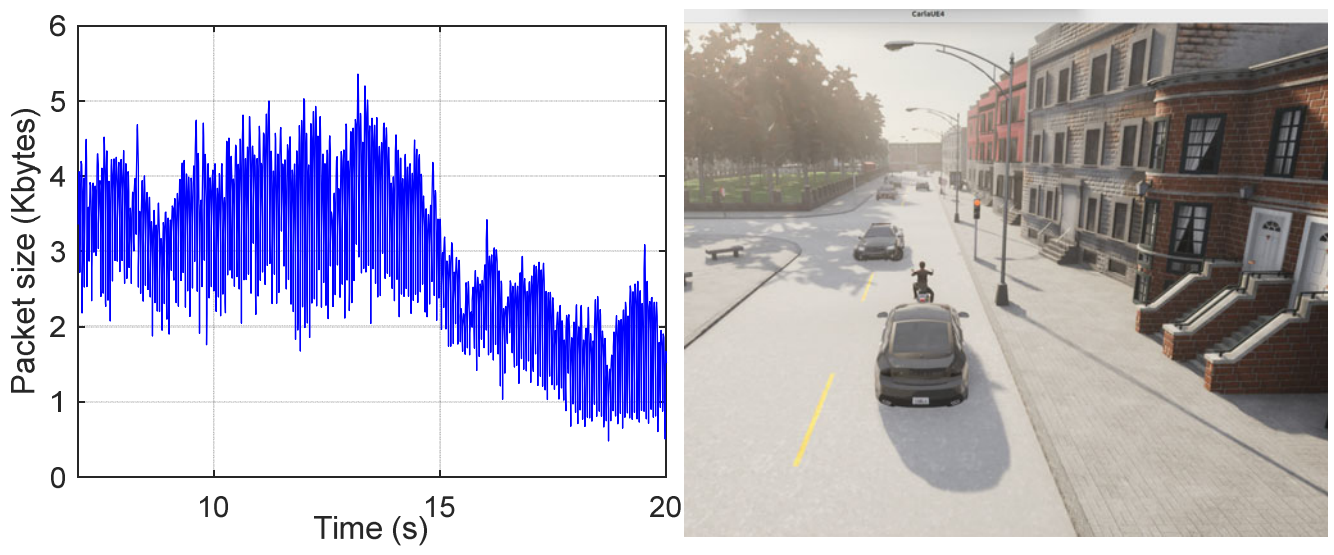

Fig. 3. Sample trace of processed sensor data packets.

Sensors (cameras and radars) generate packets at 50 ms intervals, but the object detection algorithm introduces a processing delay that depends on the driving scenario and the number of detected objects. The sensor configuration follows a cyclic pattern repeated every 50 ms, with packets from the five cameras generated first, followed by packets from the five radars. Table I reports the packet inter-arrival times between the ten packets generated every 50 ms, where $\Delta t_{ij}$ represents the time interval between packets $i$ and $j$. The first row shows the time interval between packets sequentially generated by the five cameras, while the second row reports the interval between the last camera packet and the first radar packet. Similarly, the third row details the intervals between packets from the five radars, and the fourth row shows the interval between the last radar packet and the first camera packet in the next 50 ms cycle. The table shows that the packet inter-arrival times remain nearly constant within each category, as reflected by the small standard deviation, with variations primarily due to the object detection algorithm's processing time. Given this consistency, the timing of the next ten packets can be directly estimated from the traffic characterization in Table I using the mean inter-arrival time. Importantly, any small inaccuracies in these estimations, as indicated by the standard deviation, will not impact on our predictive scheduler since it only allocates resources to the current packet. PDS uses the estimated future packet arrivals to select the radio resources for the current packet that satisfy its latency requirements and increase the likelihood of meeting the latency constraints of future packets upon their arrival.

TABLE I. PACKET INTER-ARRIVAL TIME

| Parameter | Mean (ms) | Std. dev (ms) |
|---|---|---|
| $\Delta t_{1\text{-}2}$, $\Delta t_{2\text{-}3}$, $\Delta t_{3\text{-}4}$, $\Delta t_{4\text{-}5}$ | 8.25 | 0.58 |
| $\Delta t_{5\text{-}6}$ | 3.91 | 0.62 |
| $\Delta t_{6\text{-}7}$, $\Delta t_{7\text{-}8}$, $\Delta t_{8\text{-}9}$, $\Delta t_{9\text{-}10}$ | 0.25 | 0.37 |
| $\Delta t_{10\text{-}11}$ | 13.62 | 0.89 |

The size of processed sensor packets varies significantly as shown in Fig. 3, since the size depends on the number of objects detected in the driving scenario. To predict the size of future packets, we implemented a predictor based on a Long Short-Term Memory (LSTM) network. LSTM networks are particularly suited for tasks involving sequential or time-series data, where the order and context of input elements are relevant [10]. This includes tasks where dependencies between data points over time need to be learned, making it suitable for predicting autonomous vehicle sensor data since the size of each packet is tied to past packets and sensor behaviors. The hyper-parameters used for the configuration of the LSTM are summarized in Table II and were optimized to minimize the Mean Absolute Error (MAE), which quantifies the average magnitude of prediction errors. MAE is defined in (1), where $P$ represents the number of predicted packets, $s_i$ is the actual packet size, and $\widehat{s_i}$ is the predicted packet size. We used a dataset containing 9780 samples, with 60% allocated for training the LSTM, 20% for validation, and the remaining 20% for testing.

$$MAE = \frac{1}{P}\sum_{i=1}^{P} | s_i - \widehat{s_i} | \tag{1}$$

TABLE II. LSTM HYPER-PARAMETERS

| Parameter | Value | Parameter | Value |
|---|---|---|---|
| Sequence Length | 150 | Number of layers | 3 (units: 256,128,64) |
| Batch Size | 32 | Features Per Sample | 6 |
| Dropout | 0.1 | Epochs | 100 |
| Optimizer | Adam | Scaling Method | Min-Max Scaling |
| Learning rate | 0.001 | Number of Outputs | 10 |

The LSTM network employs a regression-based supervised learning method and the predicted packet sizes can take any real numerical value. However, our analysis of the dataset revealed that packet sizes are limited to a discrete set of values that depend on the number of objects detected by the sensors and the specific sensor type. The output of the LSTM is then discretized by adjusting the predicted packet size to the nearest possible discrete packet size value.

The LSTM network is configured to predict the next ten packets, considering the previously described cyclic pattern generation. Fig. 4 illustrates the predictor's performance using the CDF of the relative absolute error for the prediction of the first, fifth and tenth packet. As expected, the figure shows that prediction accuracy decreases for later packets but remains sufficient for the purpose of the PDS. The predictor accurately forecasts 85% of the next packets without error, with this percentage decreasing to 78% and 70% for the fifth and tenth packets, respectively. The average relative error for the first prediction is only 9%, increasing to 13% and 24% for the prediction of the fifth and tenth packets. It is important to highlight that the lower prediction accuracy for later packets has a limited impact on the PDS. This is because the likelihood that the transmission windows of these distant packets overlap with the transmission window for the currently scheduled packet is low. In contrast, the higher accuracy of predictions for first packets has a more significant influence on the scheduler's performance, as their transmission windows are more likely to overlap with that of the packet being scheduled in dynamic scheduling.

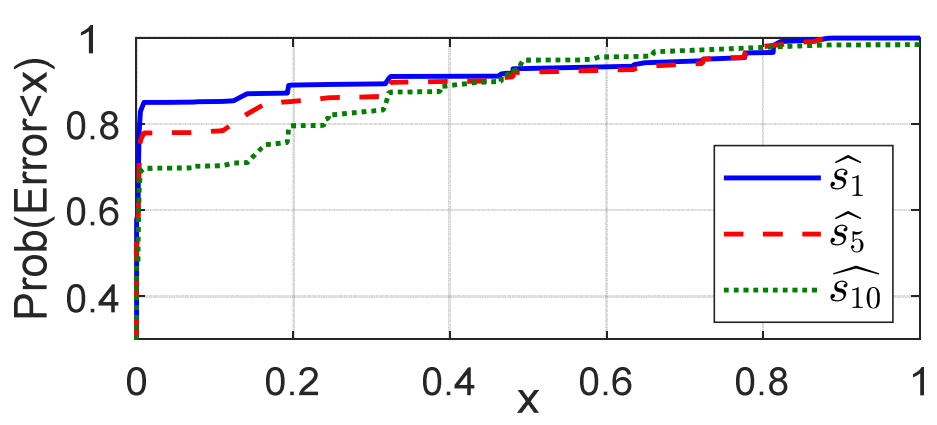

Fig. 4. CDF of the relative absolute error ($|s_i - \widehat{s_i}|/s_i$).

## V. Evaluation and Discussion

This section compares the performance of the proposed PDS scheme against a reference 5G dynamic scheduling scheme. Following [18], the reference scheme is configured to minimize the latency of each packet requesting resources, and allocates radio resources in the earliest available slot that meets the packet's transmission requirements. Unlike the PDS scheme, the reference scheduler considers only the latency requirements of the current packet and does not account for future traffic demands in its resource allocation decisions. Since each sensor generates packets every 50 ms, a packet is dropped if it cannot be scheduled for transmission before the next packet is generated 50 ms later.

Performance is evaluated in a 5G NR cell with 5 MHz bandwidth and a subcarrier spacing (SCS) of 30 kHz with a slot duration of 0.5 ms in accordance with [1]. Uplink sensor data packets are generated based on the pattern described in Section IV and transmitted using MCS6 (Modulation and Coding Scheme) to ensure their robust and reliable transmission. Each sensor data packet has a latency requirement derived from 3GPP specifications for enhanced V2X scenarios [19]. Camera-generated packets must be transmitted within a maximum latency of 50 ms. To assess different mixed traffic flow conditions, two configurations are tested for radar packets. In the first configuration (C1), the maximum latency for radar packets is randomly set to either 10 ms or 5 ms. In the second one (C2), the maximum latency for radar packets is randomly set to either 10 ms or 3 ms. We evaluate performance under different network loads, considering a single autonomous vehicle per cell (scenario S1) and two autonomous vehicles per cell (scenario S2).

Table III presents the percentage of transmissions that meet latency requirements for both the PDS scheme and the reference (Ref.) scheme across all scenarios and all latency requirement configurations. The results show that in the low-load scenario (S1) with configuration C1, the PDS scheme can support deterministic service levels as it successfully supports all transmissions within their latency deadline, whereas the reference scheme meets latency requirements for only 85.7% of them. The performance gains of the PDS scheme over standard 5G dynamic scheduling stem from its ability to anticipate short-term traffic demands and allocate current transmissions the radio resources, from all possible options, that satisfy their demand and are less likely needed by upcoming packets to satisfy their demand. By leveraging predictive knowledge of future traffic, the PDS scheme flexibly utilizes the latency budgets of mixed traffic flows with different QoS requirements to better support deterministic communications and increase the percentage of satisfied transmissions. This is visible in Fig. 5.a that plots the cumulative distribution function (CDF) of the latency experienced by packets from cameras and radars in scenario S1 under configuration C1. Table III shows that both the PDS and reference schemes satisfy all camera transmissions in S1-C1. However, Fig. 5.a reveals that while the PDS scheme ensures 100% of camera packets meet their 50 ms latency requirement, it slightly increases the latency of approximately 10% of camera packets beyond 5 ms compared to the reference scheme. This adjustment is intentional, as PDS strategically utilizes the 50 ms latency budget of camera packets to prioritize predicted radar transmissions with stricter latency constraints. As a result, Fig. 5.a shows that the PDS scheme reduces the latency of radar transmissions compared to the reference scheme, with greater improvements observed for packets with stricter latency requirements. The reference scheme lacks this predictive knowledge and meets the latency requirements for only 95% of radar transmissions with a 10 ms deadline and just 50% of radar transmissions with a 5 ms deadline (Table III), whereas PDS ensures deterministic service levels for 100% of radar transmissions.

The benefits of predictive knowledge in dynamic scheduling augments as latency requirements become stricter within mixed traffic flows. Under the C2 configuration, where certain radar packets have a latency requirement of 3 ms, Table III shows that the reference scheme satisfies only 78.6% of transmissions in S1, with this percentage dropping to just 20% for radar packets with the strictest 3 ms deadline. This is compared to 48% of satisfied radar transmissions with a 5 ms latency deadline under the C1 configuration. In contrast, the proposed PDS scheme can better support deterministic service levels as it successfully satisfies 92.9% of all transmissions and nearly 72% of radar packets with a 3 ms latency requirement. The remaining 7.1% of transmissions that fail to meet their latency constraints with PDS correspond to radar packets with large sizes that exceed the transmission capacity within the 3 ms deadline, given the available radio resources in the cell. For instance, in a best-case scenario, 66 RBs are available within a 3 ms window[1]. A radar packet containing data from five detected objects has a size of 3780 bytes and requires 67 RBs when transmitted using MCS6, making it impossible to meet the 3 ms deadline regardless of the scheduling scheme. This finding highlights that PSD effectively guarantees latency requirements for all feasible transmissions under S1-C2 considering the cell resources.

TABLE III. Percentage (%) of Transmissions Meeting Their Latency Requirements

| Latency req. | Transmissions | S1 | | S2 | |
|---|---|---|---|---|---|
| | | Ref. | PDS | Ref. | PDS |
| C1 | **Total** | **85.70** | **100** | **55.02** | **97.00** |
| | Cameras (50 ms) | 100 | 100 | 99.84 | 97.61 |
| | Radars (10 ms) | 95.35 | 100 | 12.37 | 99.89 |
| | Radars (5 ms) | 47.95 | 100 | 7.89 | 95.90 |
| C2 | **Total** | **78.60** | **92.90** | **53.95** | **87.80** |
| | Cameras (50 ms) | 100 | 100 | 99.84 | 96.00 |
| | Radars (10 ms) | 95.35 | 100 | 12.37 | 93.66 |
| | Radars (3 ms) | 20.08 | 71.93 | 3.69 | 68.95 |

When the traffic load increases (scenario S2), more packets have overlapping transmission windows (as illustrated in Fig. 2), leading to greater competition for the same radio resources. In this case, careful allocation of radio resources becomes even more critical. The gains achieved with the PDS scheme, which leverages predicted traffic demands to optimize resource allocation while accounting for both current and future (potentially conflicting) transmissions, become even more significant compared to standard 5G dynamic scheduling. Table III shows that the percentage of transmissions meeting latency requirements drops significantly for the reference scheme in S2, with only 55.02% and 53.95% of all transmissions satisfied under the C1 and C2 configurations, respectively. The performance deteriorates further for packets with the most stringent latency requirements. Specifically, Table III shows that only 7.89% of radar transmissions with a 5 ms deadline (C1) and just 3.69% of those with a 3 ms deadline (C2) are successfully transmitted with the reference scheme. In contrast, the proposed PDS scheme successfully meets the latency requirements for 97% and 87.8% of all transmissions under the higher-load scenario with mixed traffic flows, underscoring its capacity to better support deterministic service levels. Additionally, it ensures that 95.9% and 68.9% of radar transmissions with 5 ms and 3 ms latency requirements under C1 and C2, respectively, meet their deadlines. It is important to note that the PDS scheme meets the latency requirements for 88.57% of transmissions in S2-C2 using 89.8% of the available radio resources. In contrast, the reference scheme, which consumes 97.6% of the resources, manages to satisfy the latency requirements for only 54.95% of transmissions. This clearly demonstrates the most efficient resource utilization achieved by the PDS scheme by leveraging predictive knowledge of future traffic.

[1] With 5 MHz bandwidth and 30 kHz SCS, there are 11 radio resources or RB per slot, totaling 66 RB within 3 ms. This is a best-case analysis assuming that packet generation is synchronized with the start of the slot and that the scheduling requests and response do not add any additional latency.

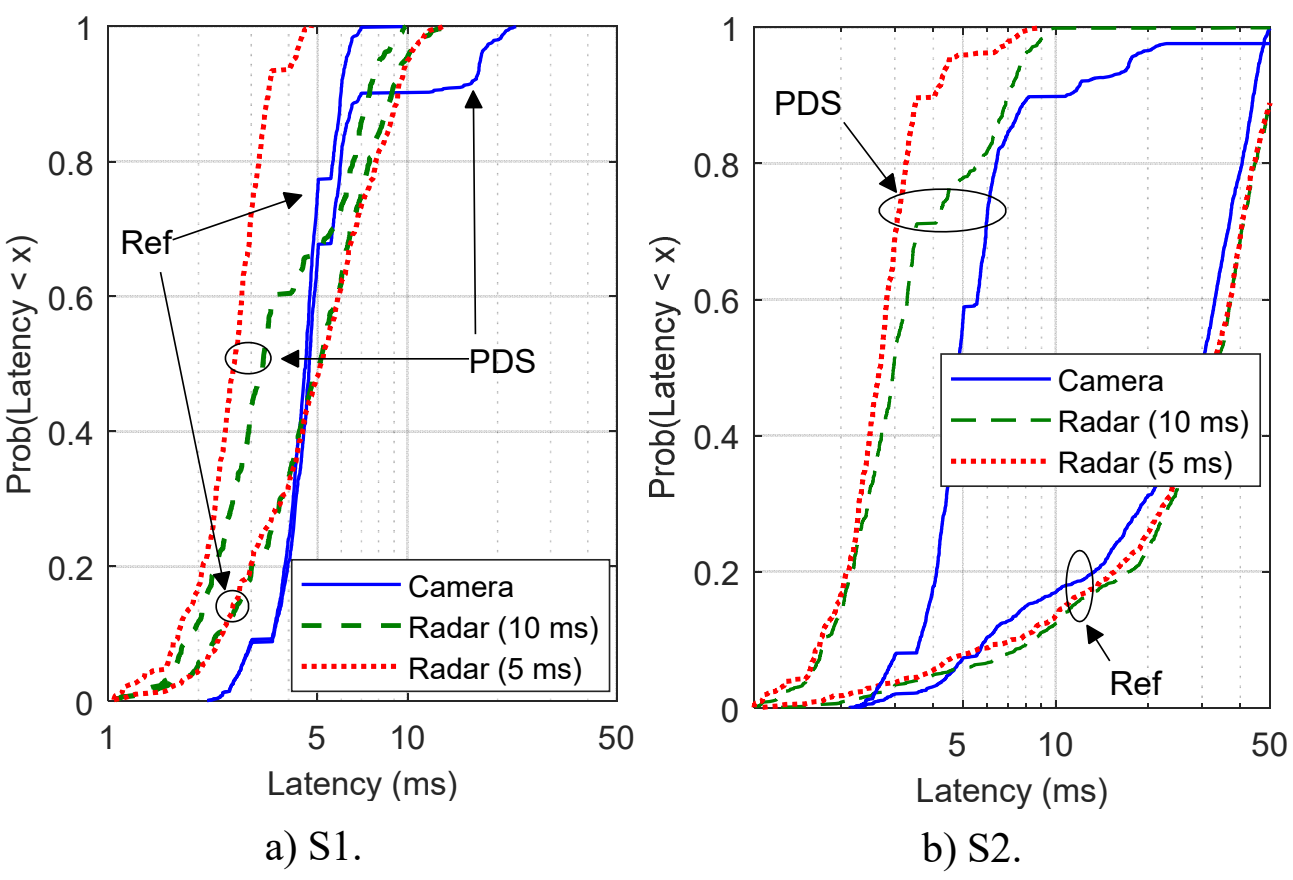

Fig. 5. CDF of the latency under C1 configuration.

The superior performance of the PDS scheme stems from its ability to anticipate short-term traffic demands and allocate radio resources to current transmissions in a way that minimizes conflicts with upcoming packets. To achieve this, the PDS scheme strategically utilizes the available latency budget of camera packets to allocate resources that ensure the timely delivery of radar packets with stricter latency constraints without compromising camera traffic with more relaxed requirements. This effect is evident in Fig. 5.b, where 10% of camera packets experience latencies exceeding 8 ms, allowing most radar transmissions to meet their deadlines. Compared to the lower-load scenario (Fig. 5.a), Fig. 5.b shows that a wider portion of the 50 ms latency budget for camera packets is leveraged to accommodate radar transmissions when the load increases. These results highlight the effectiveness of the predictive scheduler in optimizing resource allocation for mixed traffic flows with diverse QoS requirements.

## VI. Conclusions

This study presents a novel predictive dynamic scheduling scheme for beyond-5G communications. The proposed scheme leverages traffic predictions to allocate resources to incoming scheduling requests, ensuring they meet their latency requirements while avoiding resources likely to be needed by future predicted packets. Our evaluation demonstrates that the proposed scheme outperforms standard 5G dynamic scheduling and significantly enhances the ability to provide deterministic service levels with bounded latency deadlines in scenarios with mixed traffic flows and varying QoS requirements. By leveraging traffic predictions, the scheme dynamically utilizes different latency budgets to optimize resource allocations, increasing the percentage of transmissions that meet their latency constraints.

## References

[1] 3GPP, TS 38.300. V17.11.0, "NR; NR and NG-RAN Overall Description; Stage 2 (Release 18))", Release 17, 2024.

[2] Yungang Pan, et al., "Resource Optimization with 5G Configured Grant Scheduling for Real-Time Applications", in *Proc. Design, Automation & Test in Europe Conference & Exhibition (DATE)*, Antwerp, Belgium, 2023, pp. 1-2.

[3] 3GPP, TR 38.824. V16.0.0, "Study on physical layer enhancements for NR ultra-reliable and low latency case (URLLC)", Release 16, 2019.

[4] A. Larrañaga-Zumeta, M.C. Lucas-Estañ, J.Gozálvez, A. Arriola, "5G configured grant scheduling for seamless integration with TSN industrial networks", *Computer Communications*, 2024, 107930.

[5] T. Zhang, X. S. Hu, S. Han, "Contention-Free Configured Grant Scheduling for 5G URLLC Traffic", in *Proc. 60th ACM/IEEE Design Automation Conference (DAC)*, San Francisco, USA, 2023, pp. 1-6.

[6] G. P. Sharma, et al., "Toward deterministic communications in 6G networks: state of the art, open challenges and the way forward", *IEEE Access*, vol. 11, pp. 106898–106923, Sep. 2023.

[7] Q. He, G. Dán, and G. P. Koudouridis, "Semi-persistent scheduling for 5G downlink based on short-term traffic prediction", in *Proc. 2020 IEEE Global Communications Conference* (GLOBECOM 2020), Taipei, Taiwan, Dec. 2020.

[8] M. Elsayem, et al., "Intelligent Resource Allocation for Grant-Free Access: A Reinforcement Learning Approach", *IEEE Networking Letters*, vol. 5, no. 3, pp. 154-158, Sept. 2023.

[9] Y. Kaura, et al., "Adaptive Scheduling of Shared Grant-Free Resources for Heterogeneous Massive Machine type Communication in 5G and Beyond Networks", *IEEE Trans. on Network and Service Management*, Early Access, Nov. 2024.

[10] S. Cavallero, et al., "A new scheduler for URLLC in 5G NR IIoT networks with spatio-temporal traffic correlations", in *Proc. IEEE International Conference on Communications (ICC 2023)*, Rome, Italy, May 28–Jun. 1, 2023.

[11] A. Valcarce, et al., "The Role of AI in 6G MAC", in *Proc. 2024 Joint European Conference on Networks and Communications & 6G Summit (EuCNC/6G Summit)*, Antwerp, Belgium, 2024, pp. 723-728.

[12] European Technology Platform Networld Europe, "Smart Networks in the context of NGI", *Technical Annex to Strategic Research and Innovation Agenda 2022-27*, 2024.

[13] X. Hong, Y. Xi, and P. Liu, "Resource-aware online traffic scheduling for time-sensitive networking", *IEEE Transactions on Industrial Informatics*, vol. 20, no. 12, pp. 14267-14276, Dec. 2024.

[14] D. Yang, et al., "DetFed: Dynamic resource scheduling for deterministic federated learning over time-sensitive networks", *IEEE Transactions on Mobile Computing*, vol. 23, no. 5, pp. 5162-5178, May 2024.

[15] P. Prados-Garzon, et al., "LEARNET: Reinforcement learning based flow scheduling for asynchronous deterministic networks", in *Proc. 2020 IEEE International Conference on Communications (ICC 2020)*, Dublin, Ireland, Jun. 7–11, 2020.

[16] 6G IA, *European Vision for the 6G Network Ecosystem*, v2, Nov. 2024.

[17] L. Lusvarghi, et al., "Characterization of In-Vehicle Network Sensor Data Traffic in Autonomous Vehicles", in *Proc. 2024 IEEE Vehicular Networking Conference (VNC)*, Kobe, Japan, May, 2024.

[18] A. Omer, et al., "Performance Evaluation of 5G Delay-Sensitive Single-Carrier Multi-User Downlink Scheduling", in *Proc. IEEE 34th Annual International Symposium on Personal, Indoor and Mobile Radio Communications* (PIMRC), Toronto, Canada, 2023, pp. 1-6.

[19] 3GPP, TS 22.186. V17.0.0, "Enhancement of 3GPP support for V2X scenarios", Release 17, 2022.